\documentclass[11pt,a4paper]{article}
\usepackage{authblk}
\usepackage{lmodern}
\usepackage{jheppub}
\usepackage{amsmath}
\usepackage{mathtools}
\usepackage{bbold}
\usepackage{tikz}
\usepackage{empheq}
\usepackage{enumitem}
\usepackage{graphics}
\allowdisplaybreaks
\usepackage{wrapfig}
\usepackage{caption}
\usepackage{natbib}
\usepackage{xcolor}
\usepackage{framed}
\usepackage{array}
\usepackage{dsfont}
\usepackage{braket}
\definecolor{shadecolor}{gray}{0.925}
\usepackage[cbgreek]{textgreek}

\def\sideremark#1{\ifvmode\leavevmode\fi\vadjust{\vbox to0pt{\vss
 \hbox to 0pt{\hskip\hsize\hskip1em
 \vbox{\hsize3cm\tiny\raggedright\pretolerance10000
 \noindent #1\hfill}\hss}\vbox to8pt{\vfil}\vss}}}%

\newcommand{\bi}{\begin{itemize}}
\newcommand{\ei}{\end{itemize}}
\newcommand{\bea}{\begin{align}}
\newcommand{\eea}{\end{align}}
\newcommand{\be}{\begin{equation}}
\newcommand{\ee}{\end{equation}}

\newcommand{\eq}{Eq.~}
\newcommand{\eqs}{Eqs.~}

\makeatletter
\renewcommand*\env@matrix[1][\arraystretch]{%
  \edef\arraystretch{#1}%
  \hskip -\arraycolsep
  \let\@ifnextchar\new@ifnextchar
  \array{*\c@MaxMatrixCols c}}
\makeatother

\author[\ensuremath{a},\ensuremath{b}]{Renjie Song}
\author[\ensuremath{a}]{\quad  Mingchen Zheng}
\author[\ensuremath{a},\ensuremath{b},\ensuremath{c}]{\quad Junpeng Cao}
\author[\ensuremath{a}]{\quad Yupeng Wang}

\affiliation[\ensuremath{a}]{Beijing National Laboratory for Condensed Matter Physics, Institute of
Physics, Chinese Academy of Sciences, Beijing 100190, China}

\affiliation[\ensuremath{b}]{School of Physical Sciences, University of Chinese Academy of Sciences,
Beijing 100049, China}

\affiliation[\ensuremath{c}]{Peng Huanwu Center for Fundamental Theory, Xi'an 710127, China}

\emailAdd{txsong@iphy.ac.cn
zhengmc@iphy.ac.cn, 
junpengcao@iphy.ac.cn, 
yupeng@iphy.ac.cn}

\title{\centering \huge An integrable Anderson-impurity problem embedded in the one-dimensional Hubbard model}

\abstract{ An exactly solvable one-dimensional Hubbard model with a single Anderson impurity embedded at the boundary is constructed in the framework of the quantum inverse scattering method. The model is solved exactly by the nested Bethe ansatz method. We identify the boundary bound states and determine the ground state phase diagram. By deriving the impurity contribution to the magnetic susceptibility, we show that in the dilute electron limit, a nearly free local moment forms at the impurity site, while at finite electron densities, the impurity spin is screened by the host electrons, consistent with Kondo physics.}

\begin{document}

\begin{flushright}    
\texttt{}
\end{flushright}

\maketitle 

\newpage

\section{Introduction}

Impurity systems provide a sensitive probe of strongly correlated quantum matter, where a local perturbation can induce significant changes in global properties \cite{Kondo1970, Hewson1997}. A paradigmatic example is the Anderson model \cite{Anderson1961}, whose Hamiltonian is given by
\begin{equation}
    H_{imp} = \sum_{k\sigma} \epsilon_k\, c_{k\sigma}^\dagger c_{k\sigma}
    + \sum_{k\sigma} \left( V_k\, c_{k\sigma}^\dagger d_\sigma + \text{H.c.} \right) + \sum_{\sigma} \epsilon_d\, d_{\sigma}^\dagger d_{\sigma}
    + \mathcal{U}\, n_{d\uparrow} n_{d\downarrow},
\end{equation}
where $c_{k\sigma}^\dagger$ creates a conduction electron with momentum $k$ and spin $\sigma$, and $d_{\sigma}^\dagger$ creates an electron on the localized impurity level with energy $\epsilon_d$. $\mathcal{U}\, n_{d\uparrow} n_{d\downarrow}$ represents the on-site Coulomb repulsion at the impurity, and the hybridization $V_k$ describes tunneling between the impurity and the conduction band. This model captures the interplay between impurity–bath hybridization and strong local Coulomb interactions, serving as a cornerstone for the theoretical study of quantum impurity effects. 
These include the formation of local magnetic moments \cite{Anderson1961}, the Kondo effect \cite{Kondo1964, Schrieffer1966}, Coulomb blockade in mesoscopic systems \cite{Beenakker1991, Goldhaber1998} and the emergence of heavy fermion physics in correlated materials \cite{Hewson1997}. 

Over the decades, the Anderson model has been extensively studied by means of a variety of analytical and numerical methods \cite{wilson1975, Andrei2, Wiegmann3}. Notably, since only the $s$-wave channel couples to the impurity, this model effectively reduces to one dimension and becomes integrable in the isotropic limit \cite{Andrei2, Wiegmann3, Wiegmann1,Wiegmann2}. This integrability enables exact solutions via the Bethe ansatz, providing rigorous, non-perturbative insights into impurity behavior. 
An important feature of the Anderson model is that its conduction bath is non-interacting, with hybridizing electrons treated as free fermions. It is then natural to ask whether Anderson-type impurities can induce similarly rich and nontrivial phenomena when embedded in a host with interactions. 

The Hubbard model provides a controlled framework for the exact study of impurity phenomena in a strongly correlated host \cite{Hubbard_book}.
Its integrability in one dimension was first found by Lieb and Wu via the coordinate Bethe ansatz \cite{Lieb1}, and subsequently reformulated in terms of the Yang--Baxter structure by Shastry \cite{Shastry1, Shastry2}. Building on this framework, the quantum inverse scattering method \cite{Takhtadzhan1979QISM, Sklyanin1979QISM} provides a systematic approach for constructing integrable impurity models by suitably modifying the local structure of the $L$-operators. Using this method, a variety of exactly solvable impurity models have been constructed, including the Kondo impurity in the Heisenberg spin chain \cite{Andrei1}, integrable spin chains with alternating or higher spins \cite{Schlottmann1, Schlottmann2, DeVega1, DeVega2}, and $t$--$J$ models with localized spin impurities \cite{Wang1997}. 
While integrable extensions of the Hubbard model with open boundaries, such as those incorporating boundary chemical potentials or magnetic fields, have been extensively studied \cite{Wang1996, Asakawa1, Asakawa4, Asakawa5, Wadati2, Deguchi1}, integrable realizations of Anderson-type magnetic impurities within the Hubbard chain remain largely unexplored. 

In this work, within the framework of the quantum inverse scattering method, we construct a new integrable one-dimensional Hubbard model with a single Anderson impurity at the boundary. This model is exactly solved using the nested Bethe ansatz method, which yields a set of Bethe ansatz equations (BAEs). By analyzing the BAEs, we identify boundary bound states in different parameter regimes and obtain the corresponding impurity phase diagram. We also derive expressions for impurity-induced charge and spin densities and, employing the Wiener–Hopf method, compute the impurity contribution to the magnetic susceptibility. Our analysis reveals that in the dilute limit the impurity behaves as a nearly free local moment, while at finite electron densities, the finiteness of the susceptibility indicates the Kondo screening as in the Kondo problem. These results provide a detailed characterization of impurity magnetic properties in a strongly correlated host. 

The paper is organized as follows. In Section \ref{sec:construction model}, we construct the model Hamiltonian via the quantum inverse scattering method. Section \ref{sec:BA solution} presents its exact solution using the nested Bethe ansatz. In Section \ref{sec:Bound state}, we analyze the BAEs to identify boundary bound states and characterize the associated ground state phases. Section \ref{sec:continuum limit} derives coupled integral equations for the charge and spin densities in the thermodynamic limit. Section \ref{sec:susceptibility} focuses on the impurity contribution to the magnetic susceptibility. Finally, Section \ref{conclusion} concludes the paper.

\section{An integrable Anderson impurity model}
\label{sec:construction model}

The Hamiltonian of the one-dimensional Hubbard model with open boundary condition is
\begin{equation}
    H_0 = - \sum_{j=1}^{L-1} \sum_\sigma \left( c_{j\sigma}^\dagger c_{j+1\sigma}+ c_{j+1\sigma}^\dagger c_{j\sigma} \right) 
  + U \sum_{j=1}^L n_{j\uparrow}n_{j\downarrow} ,
\end{equation} 
where $L$ is the lattice number, $c_{j \sigma}^{\dagger}$ and $c_{j \sigma}$ denote the fermionic creation and annihilation operators with spin $\sigma$ ($\uparrow $ or $\downarrow$) at site $j$, $n_{j\sigma}=c_{j \sigma}^{\dagger}c_{j \sigma}$ is the number operator, and $U$ denotes the on-site Coulomb interaction strength. 

\vskip 4pt

The integrable structure of the model is encoded in the $R$-matrix formalism, which governs the scattering processes. 
The $R$-matrix of the Hubbard model depends on two spectral parameters $\theta_1$ and $\theta_2$, and takes the form \cite{Shastry3, Wadati1, Wadati3}:
\begin{equation}\label{R_Matrix} 
\begin{split}
    R_{1,2}(\theta_1, \theta_2) = &\cosh(\tilde{h}_1 - \tilde{h}_2) 
    L^{(\sigma)}_{12}(\theta_1 - \theta_2) \otimes L^{(\tau)}_{12}(\theta_1 - \theta_2) \\
    &+ \frac{\cos(\theta_1 - \theta_2)}{\cos(\theta_1 + \theta_2)} \sinh(\tilde{h}_1 - \tilde{h}_2) 
    L^{(\sigma)}_{12}(\theta_1 + \theta_2) \otimes L^{(\tau)}_{12}(\theta_1 + \theta_2) \sigma_1^z \tau_1^z,
\end{split} 
\end{equation}
where the subscripts $\{1,2\}$ imply that $R$-matrix acts in the tensor space $V_1 \otimes V_2$ and $\tilde{h}_i \equiv \tilde{h}(\theta_i)$ is a nonlinear function of the spectral parameter $\theta_i$, defined by
\begin{equation}\label{sinh2h}
    \frac{\sinh 2\tilde{h}(\theta_i)}{\sin 2\theta_i} = \frac{U}{4}. 
\end{equation}
The operators $L^{(\sigma)}_{12}(\theta)$ and $L^{(\tau)}_{12}(\theta)$ act on two independent spin-$\tfrac{1}{2}$ auxiliary spaces and are given by
\begin{align}
    L^{(\sigma)}_{12}(\theta) 
    &=\frac{\cos\theta +\sin\theta}{2} + \frac{\cos\theta - \sin\theta}{2} \sigma_1^z \sigma_2^z
    + \sigma_1^+ \sigma_2^- + \sigma_1^- \sigma_2^+, \\
    L^{(\tau)}_{12}(\theta) 
    &=\frac{\cos\theta +\sin\theta}{2} + \frac{\cos\theta - \sin\theta}{2} \tau_1^z \tau_2^z
    + \tau_1^+ \tau_2^- + \tau_1^- \tau_2^+,
\end{align}
where $\{\sigma^\pm$, $\sigma^z\}$ and $\{\tau^\pm$, $\tau^z\}$ are two sets of standard Pauli operators. 
It is shown that the $R$-matrix \eqref{R_Matrix} satisfies the Yang-Baxter equation \cite{Wadati1} 
\begin{equation}
    R_{1,2}(\theta _1,\theta _2)R_{1,3}(\theta _1,\theta _3)R_{2,3}(\theta _2,\theta _3)
    =R_{2,3}(\theta _2,\theta _3)R_{1,3}(\theta _1,\theta _3)R_{1,2}(\theta _1,\theta _2).
\end{equation}

\vskip 4pt

To incorporate a boundary impurity while preserving integrability, we adopt the reflection algebra framework and introduce boundary reflection matrices $K^\pm(\theta)$ \cite{Sklyanin1988}. In this work, we consider diagonal $K$-matrices of the form \cite{Zhou1, Wadati3}
\begin{align} 
    K^+(\theta) &= \mathbb{1} \otimes \mathbb{1}\label{Kplus}, \\
    K^-(\theta) &= \mathbb{1} \otimes \mathbb{1} 
    - p (\sigma^z \otimes \mathbb{1} + \mathbb{1} \otimes \sigma^z) \theta\label{Kminus}, 
\end{align}
where $p$ is a boundary coupling parameter that characterizes the impurity strength. The $R$-matrix~\eqref{R_Matrix} and $K$-matrices  satisfy the reflection equations \cite{Zhou1}
\begin{align}
    &R_{1,2}(\theta_1,\theta_2) K_1^-(\theta_1) R_{2,1}(\theta_2,-\theta_1) K_2^-(\theta_2) \notag \\
    = &K_2^-(\theta_2) R_{1,2}(\theta_1,-\theta_2) K_1^-(\theta_1) 
    R_{2,1}(-\theta_2,-\theta_1)\label{RE1}, \\
    &R_{2,1}(\theta_2,\theta_1) K_1^{+t_1}(\theta_1) R_{1,2}(-\theta_1-\pi,\theta_2) 
    K_2^{+t_2}(\theta_2) \notag \\
    = &K_2^{+t_2}(\theta_2) R_{2,1}(-\theta_2-\pi,\theta_1) K_1^{+t_1}(\theta_1) 
    R_{1,2}(-\theta_1,-\theta_2)\label{RE2}.
\end{align}
The one-row monodromy matrices $T_0(\theta,\varphi)$, $\hat{T}_0(\theta,\varphi)$ and the double-row monodromy matrix $\mathcal{U}_0(\theta)$ are defined as follows
\begin{align}
    T_0(\theta,\varphi)
    &=R_{0,L}(\theta,0) \dots R_{0,2}(\theta,0)R_{0,1}(\theta,\varphi), \\
    \hat{T}_0(\theta,\varphi)
    &=\varrho(\theta,\varphi) T_0^{-1}(-\theta,\varphi),\\
    \mathcal{U}_0(\theta,\varphi)& = T_0(\theta,\varphi) K_0^-(\theta) \hat{T}_0(\theta,\varphi),  
\end{align}
where $0$ represents the auxiliary space and $\varrho(\theta,\varphi) = \rho(-\theta,\varphi) \rho^{L-1}(-\theta,0)$, here
\begin{equation}
    \rho(\theta_1,\theta_2) 
    = R_{1,2}(\theta_1,\theta_2) R_{2,1}(\theta_2,\theta_1). 
\end{equation}
The transfer matrix is given by 
\begin{equation}\label{transfer_matrix1}
    t(\theta,\varphi) = tr_0 \left[ K_0^+(\theta)\, \mathcal{U}_0(\theta,\varphi) \right].
\end{equation}
From \eqref{RE1} and \eqref{RE2}, one obtains the RTT relation for the monodromy matrix,
\begin{equation}
R_{1,2}(\theta_1,\theta_2), T_1(\theta_1,\varphi), T_2(\theta_2,\varphi)
= T_2(\theta_2,\varphi), T_1(\theta_1,\varphi), R_{1,2}(\theta_1,\theta_2), 
\end{equation}
Thus the transfer matrices~\eqref{transfer_matrix1} commute for different spectral parameters,
\begin{equation}
[t(\theta_1,\varphi), t(\theta_2,\varphi)] = 0,
\end{equation}
which generates an infinite family of conserved quantities.

\vskip 4pt

The Hamiltonian of the integrable Hubbard chain with boundary impurity is then derived from the derivative of the transfer matrix~\eqref{transfer_matrix1} at $\theta = 0$, and reads
\begin{equation}\label{Hbar}
\begin{split}
    H_{\theta=0} = &\frac{1}{8 \rho(0,\varphi)} 
    \left. \frac{\partial t(\theta,\varphi)}{\partial \theta} \right|_{\theta=0} \\
    = &\sum_{j=2}^{L-1} 
    \left[ (\sigma_j^+ \sigma_{j+1}^- + \sigma_j^- \sigma_{j+1}^+) 
    + (\tau_j^+ \tau_{j+1}^- + \tau_j^- \tau_{j+1}^+) \right] 
    + \frac{U}{4} \sum_{j=1}^L \sigma_j^z \tau_j^z \\
    &+\left[ 
    \left( \frac{1}{\cos\varphi} - p \frac{\sin\varphi}{\cos^2\varphi} \cosh2\tilde{h} \right)(\sigma_1^+ \sigma_2^- + \tau_1^+ \tau_2^-) \right.\\
    &+ \left. \left( \frac{1}{\cos\varphi} + p \frac{\sin\varphi}{\cos^2\varphi} \cosh2\tilde{h} \right)(\sigma_1^- \sigma_2^+ + \tau_1^- \tau_2^+)\right] \\
    &+ p \frac{\sin\varphi}{\cos^2\varphi} \sinh2\tilde{h} 
    \left[ \tau_1^z (\sigma_1^+ \sigma_2^- + \sigma_1^- \sigma_2^+) 
    + \sigma_1^z (\tau_1^+ \tau_2^- + \tau_1^- \tau_2^+) \right] \\
    &+ \tan\varphi \sinh 2\tilde{h} \sigma_1^z \tau_1^z - \frac p2 \frac{1}{\cos^2\varphi} (\sigma_1^z + \tau_1^z) 
    +\frac p2 \tan^2\varphi (\sigma_2^z + \tau_2^z) + \mathrm{const}.
\end{split}
\end{equation}
The function $\tilde{h}(=\tilde{h}(\varphi))$ is defined in~\eqref{sinh2h}, and the parameter $\varphi$ is taken to be purely imaginary to ensure the Hermiticity of the Hamiltonian. Using the Jordan--Wigner transformation
\begin{equation}
    \begin{split}
    c_{j\uparrow} &= ( \sigma_1^z \dots \sigma_{j-1}^z ) \sigma_j^-, \\  
    c_{j\downarrow} &= ( \sigma_1^z \dots \sigma_{L}^z ) 
    ( \tau_1^z \dots \tau_{j-1}^z ) \tau_j^-,
    \end{split}
\end{equation}  
the Hamiltonian $H_{\theta=0}$~\eqref{Hbar} becomes 
\begin{equation}\label{Hamiltonian_1}
\begin{split}
  H = &- \sum_{j=2}^{L-1} \sum_\sigma \left( c_{j\sigma}^\dagger c_{j+1\sigma}+ c_{j+1\sigma}^\dagger c_{j\sigma} \right) 
  + U \sum_{j=2}^L n_{j\uparrow}n_{j\downarrow} \\
  &+ \varepsilon_1 n_1 + \gamma U n_{1\uparrow} n_{1\downarrow} 
  + \sum_\sigma 
  \left( V c_{1\sigma}^\dagger c_{2\sigma} + V^* c_{2\sigma}^\dagger c_{1\sigma} \right) 
    \\
  &+ \varepsilon_2 n_2 + \Delta \sum_\sigma n_{1 \bar\sigma} \left( c_{1\sigma}^\dagger c_{2\sigma} + c_{2\sigma}^\dagger c_{1\sigma} \right),
\end{split}
\end{equation}
where the constant terms are abandoned and
\begin{equation}\label{parameter_relation1}
\begin{split}  
    V &= - \frac{1}{\cos\varphi} + p \frac{\sin\varphi}{\cos^2\varphi} \cosh 2\tilde{h}(\varphi) 
    + p \frac{U}{2}\frac{\sin^2\varphi}{\cos\varphi},\quad
    \gamma = 1 + 2 \sin^2\varphi,\\
    \varepsilon_1 &= - U \sin^2\varphi - p \frac{1}{\cos^2\varphi},\quad
    \varepsilon_2 = p \tan^2\varphi, \quad
    \Delta = - pU \frac{\sin^2\varphi}{\cos\varphi}.
\end{split}
\end{equation}
Using \eq\eqref{sinh2h} and defining $\phi=\text{Im}(\varphi)$, relations~\eqref{parameter_relation1} can be rewritten as
\begin{equation}
\begin{split}    
    V &= - \frac{1}{\cosh \phi} 
    - p \frac{U}{2}\frac{\sinh^2 \phi}{\cosh \phi}
    + i \, p \frac{\sinh\phi}{\cosh^2\phi} 
    \sqrt{1 - \frac{U^2}{16} (\sinh2\phi)^2 }, \quad
    \gamma = 1 - 2 \sinh^2 \phi, \\
    \varepsilon_1 &= U \sinh^2 \phi - p \frac{1}{\cosh^2 \phi}, \quad 
    \varepsilon_2 = - p \tanh^2 \phi, \quad
    \Delta = pU \frac{\sinh^2 \phi}{\cosh \phi}.
\end{split}
\end{equation}
The Hamiltonian~\eqref{Hamiltonian_1} describes a strongly correlated one-dimensional Hubbard chain with a boundary Anderson impurity at site $j=1$. This impurity is incorporated via extra boundary terms: the on-site potential $\varepsilon_1 n_1$ and interaction $\gamma U n_{1\uparrow} n_{1\downarrow}$ set the local energy and Coulomb interaction at the impurity, while the hybridization term $V c_{1\sigma}^\dagger c_{2\sigma} + V^* c_{2\sigma}^\dagger c_{1\sigma}$ together with the spin-dependent hopping term $\Delta n_{1\bar\sigma} (c_{1\sigma}^\dagger c_{2\sigma} + c_{2\sigma}^\dagger c_{1\sigma})$ describes the coupling between the impurity and its neighboring bulk site. The adjacent site potential $\varepsilon_2 n_2$ further modifies the local energy near the impurity. All these boundary parameters are determined by the integrable model parameters $p$, $U$, and $\phi = \text{Im}(\varphi)$. 
The parameter regime of interest is characterized by repulsive interaction $U>0$ 
and $\gamma>0$, which gives
\begin{equation}
    |\phi| < \operatorname{arcsinh}\!\left(\tfrac{1}{\sqrt{2}}\right).
\end{equation}
In addition, we restrict to attractive impurity levels $\varepsilon_1$ and $\varepsilon_2$ corresponding to attractive potentials,
i.e., $\varepsilon_1,\varepsilon_2<0$, which implies 
\begin{equation}\label{constraint p}
p > \frac{U}{4}\sinh^2(2\phi).
\end{equation}

It is noted that $\phi=0$ corresponds to an open Hubbard chain with only boundary potential. In this case, the Hamiltonian~\eqref{Hamiltonian_1} is simplified as
\begin{equation}
    H_{\phi=0} = - \sum_{j=1}^{L-1} \sum_\sigma \left( c_{j\sigma}^\dagger c_{j+1\sigma}+ c_{j+1\sigma}^\dagger c_{j\sigma} \right) 
  + U \sum_{j=1}^L n_{j\uparrow}n_{j\downarrow} 
  - p (n_{1\uparrow} + n_{1\downarrow}).
\end{equation}
This model has been extensively investigated in the literature \cite{Asakawa1, Asakawa4, Asakawa5, Wadati2}. 

To investigate the response of the system to external fields, we introduce a chemical potential $\mu$ and a uniform magnetic field $h$ into the Hamiltonian~\eqref{Hamiltonian_1}, leading to the modified Hamiltonian
\begin{equation}\label{Hamiltonian_2}
    H_{\mu,h} = H + \mu \sum_{j=1}^L (n_{j\uparrow} + n_{j\downarrow}) 
    - \frac{h}{2} \sum_{j=1}^L (n_{j\uparrow} - n_{j\downarrow}).
\end{equation}
Since the particle number and the total magnetization are conserved quantities of the integrable Hubbard model, the additional terms in $H_{\mu,h}$~\eqref{Hamiltonian_2} commute with the Hamiltonian ~\eqref{Hamiltonian_1}. Thus, the inclusion of $\mu$ and $h$ preserves integrability and enables an exact treatment of impurity-induced magnetic properties in external fields.

\section{Exact Solution}
\label{sec:BA solution}
In this section, we use the nested Bethe ansatz method to solve the eigenvalue problem of the impurity model~\eqref{Hamiltonian_2}.
Since the particle number is a conserved quantity, the eigenstates of the model can be constructed as \cite{Chen1998}
\begin{equation}\label{wave_function}
\begin{split}
    \ket{\Psi} = &\sum_{j=1}^N \sum_{x_j \neq 1}^L \sum_{\alpha} 
    f_0^{\{ \alpha \}} (x_1,\dots,x_N) c_{x_1,\alpha_1}^\dagger \dots c_{x_N,\alpha_N}^\dagger\ket{0} \\
    &+ \sum_{j=2}^{N} \sum_{x_j \neq 1}^L \sum_{\alpha} 
    f_1^{\{ \alpha \}} (x_2,\dots,x_N) c_{1,\alpha_1}^\dagger c_{x_2,\alpha_2}^\dagger \dots c_{x_{N},\alpha_N}^\dagger\ket{0} \\
    &+ \sum_{j=3}^{N} \sum_{x_j \neq 1}^L \sum_{\alpha} 
    f_2^{\{ \alpha \}} (x_3,\dots,x_N) c_{1,\alpha_1}^\dagger c_{1,\alpha_2}^\dagger c_{x_3,\alpha_3}^\dagger \dots c_{x_N,\alpha_N}^\dagger\ket{0},
\end{split}
\end{equation}
where the integer $N$ is the number of electrons, $\{\alpha\}=\{\alpha_1,\dots,\alpha_N\}$ with $\alpha_j\in\{\uparrow,\downarrow\}$, $\ket{0}$ is the vacuum state, and $f_m^{\{\alpha\}}$ ($m=0,1,2$) are the corresponding wave functions for having $m$ electrons on the first lattice site.
In this paper, we focus on the case  
$1\leq N\leq L$.
Here we restrict our consider to the case of at most half-filled.
For the case $x_j\in[2,L]$, we take the wave function $f_0^{\{ \alpha \}} (x_1,\dots,x_N)$ 
as the Bethe ansatz type
\begin{equation}\label{g_wave_function}
    f_0^{\{ \alpha \}} (x_1,\dots,x_N) 
    = \sum_{p,q,r} A_p^{\{ \alpha \},r} (q) 
    \exp\left( i\sum_{j=1}^N r_{p_j} k_{p_j}x_{q_j} \right) 
    \theta(x_{q_1} \leq x_{q_2} \leq \dots \leq x_{q_N}), 
\end{equation}
where $q=\{q_1,\dots,q_N\}$ are the permutations of $\{1,\dots,N\}$; $r=\{r_1,\dots,r_N\}$ with $r_j=\pm$, $p=\{p_1,\dots,p_N\}$
and $\theta(x_{q_1}\leq x_{q_2}\leq\dots\leq x_{q_N})$ is the generalized step function, which equals one in the noted variables’ region and zero otherwise.
Using the ansatz \eqref{g_wave_function},
the functions
$f_1^{\{ \alpha \}} (x_2,\dots,x_N)$ 
and 
$f_2^{\{ \alpha \}} (x_3,\dots,x_N)$ 
can be obtained by 
$f_0^{\{ \alpha \}} (x_1,\dots,x_N)$
consistently through the eigenvalue equations. 
For all $x_j\ne1,L$ and $x_j\ne x_l$ cases, the eigenvalue equation $H\ket{\Psi}=E\ket{\Psi}$ gives
\begin{equation}\label{energy_eigenvalue}
     E = -2 \sum_{j=1}^N \cos k_j + \mu N - \frac{h}{2} \left( N - 2M \right). 
\end{equation}
For two particles occupying the same site case and apart from the boundary, by substituting the ansatz \eqref{g_wave_function} into the eigenvalue equation, we find that the wave function amplitudes satisfy the relation
\begin{equation}
    A_p^{r}(q)=S_{q_j,q_{j+1}}(r_{p_j}k_{p_j},r_{p_{j+1}}k_{p_{j+1}})A_{p'}^{r'}(q'), \label{S_Matrix}
\end{equation}
with $r'=\{\dots,r_{p_{j+1}},r_{p_j},\dots,\}$, 
$p'=\{\dots,p_{j+1},p_{j},\dots \}$ and $q'=\{\dots,q_{j+1},q_j,\dots\}$. 
For convenience, the superscript $\{\alpha\}$ is omitted and $A_p^{r}(q)$ is represented as a column vector in spin space.
The $S$-matrix is
\begin{equation}
    S_{m,n}(r_j k_j, r_l k_l)
    = \frac{\sin (r_j k_j) - \sin (r_l k_l) - i\frac{U}{2}\,P_{mn}}
           {\sin (r_j k_j) - \sin (r_l k_l) - i\frac{U}{2}},
\end{equation}
where $P_{mn}$ is the spin-exchange operator acting on the two-particle spin space
\begin{equation}
    P_{mn}A^{\{ \dots \alpha_{m} \dots \alpha_{n} \dots\},r}_p(q)
    = A^{\{ \dots \alpha_{n} \dots \alpha_{m} \dots\},r}_p(q).
\end{equation} 
Considering the case where $x_j=1$ and $x_l\in[3,L-1]$, we obtain
\begin{equation}
    A^{(+,\,\dots)}_p(q) = \bar{K}^+(k_{p_1}) A^{(-,\,\dots)}_p(q),
\end{equation}
where $\bar{K}^+(k_j)$ is a scalar function and reads
\begin{equation}
    \bar K^+ (k) 
    = - \frac
    {e^{-i2k} \left( |V|^2 - \varepsilon_1\varepsilon_2 - 2 \varepsilon_2 \cos k \right) 
    - e^{-ik} (\varepsilon_1 + 2 \cos k)}
    {e^{i2k} \left( |V|^2 - \varepsilon_1\varepsilon_2 - 2 \varepsilon_2 \cos k \right) 
    - e^{ik} (\varepsilon_1 + 2 \cos k)}.
\end{equation}
Similarly, for the case of $x_l=L$ and $x_j\in[3,L-1]$, we have
\begin{equation}
    e^{-2ik_{P_N}L} A^{(\dots\,,-)}_p(q) 
    = \bar{K}^-(k_{P_N}) A^{(\dots\,,+)}_p(q),
\end{equation}
with 
\begin{equation}
    \bar{K}^-(k) = - e^{2i k}.
\end{equation}
We have verified that the ansatz \eqref{wave_function} is also valid for the remaining cases of $x_j$ and $x_l$ when the $S$-matrix and $K^\pm$ take the above forms.

Consider the following sequence: The $j$-th particle scatters successively with all particles to its left, and is reflected at the left boundary. After this reflection it traverses the chain, scatters with the particles it encounters, is reflected at the right boundary, and returns to its original site. The process is described by the relations
\begin{equation}\label{scattering_equation}
\begin{split}
    e^{-2ik_j L} A^{(\dots,+,\dots)} 
    = &S_{j-1,j}(k_{j-1},k_j) \dots S_{1,j}(k_1,k_j) \bar K^+(k_j) S_{j,1}(-k_j,k_1) \dots \\
    &\times S_{j,j-1}(-k_j,k_{j-1}) S_{j,j+1}(-k_j,k_{j+1}) \dots S_{j,N}(-k_j,k_N) \\
    &\times \bar K^-(k_j) S_{N,j}(k_N,k_j) \dots S_{j+1,j}(k_{j+1},k_j)
    A^{(\dots,+,\dots)} .
\end{split}
\end{equation}
Using the algebraic Bethe ansatz, we solve the eigenvalue problem \eqref{scattering_equation}, and obtain the BAEs 
\begin{equation} \label{BAEs}
    \begin{split}
        &e^{i k_j 2(L+1)} Z(k_j) 
        = \prod_{\alpha=1}^M 
        \frac{\sin k_j + \lambda_\alpha + iu}{\sin k_j + \lambda_\alpha - iu} 
        \frac{\sin k_j - \lambda_\alpha + iu}{\sin k_j - \lambda_\alpha - iu},\ \ j=1,...,N,\\
        &\prod_{j=1}^N 
        \frac{\lambda_\alpha - \sin k_j + iu}{\lambda_\alpha - \sin k_j - iu} \frac{\lambda_\alpha + \sin k_j + iu}{\lambda_\alpha + \sin k_j - iu}\\
        &\hspace{6.5em} =\prod_{\beta\neq\alpha}^M
        \frac{\lambda_\alpha - \lambda_\beta + i2u}{\lambda_\alpha - \lambda_\beta - i2u}
        \frac{\lambda_\alpha + \lambda_\beta + i2u}{\lambda_\alpha + \lambda_\beta - i2u},\ \ \alpha=1,...,M,
    \end{split}
\end{equation}
where $u=U/4$, $\lambda_\alpha s$ are the introduced spin rapidities, $M$ is the number of down-spins, and $Z(k)=-\bar K^+(k)$.
\eqs\eqref{energy_eigenvalue} and~\eqref{BAEs} fully determine the spectrum of the Hamiltonian~\eqref{Hamiltonian_2}. 
For small-size systems, we verified that the spectrum obtained from the BAEs~\eqref{BAEs} is consistent with the results of exact diagonalization. 

\section{Boundary bound states}
\label{sec:Bound state}
In both the periodic and open-boundary Hubbard models, the ground state solutions of the BAEs generally consist of real Bethe roots ($\{k_j\}$ and $\{\lambda_\alpha\}$) in the thermodynamic limit. 
However, introducing an impurity can induce boundary string solutions \cite{Skorik1995BBS, Kapustin1996BBS, Tsuchiya1997BBS, Bedurftig1997, Deguchi2, Alba2013BBS, Frahm2006}, in which some Bethe roots become purely imaginary, reflecting the localized effect of the impurity. 
In our model, when the impurity effect is weak (small $p$ and $\phi$), the ground state still corresponds to all-real Bethe roots. 
As the impurity effect increases, boundary string solutions appear with lower energy than the all-real solution, indicating that the ground state becomes a boundary bound state. 
Through the analysis of the BAEs~\eqref{BAEs}, the exact criteria for the appearance of boundary string solutions can be established.

Assuming that there is a solution of BAEs~\eqref{BAEs} that consists of $N-1$ real Bethe roots $k_1,...,k_{N-1}$, $M$ real $\lambda_\alpha$s and a single purely imaginary Bethe root $k_N=i\kappa$ \footnote{We begin by assuming $k_N$ is purely imaginary. In fact, if one instead allows for a finite real part, the subsequent analysis shows that this real part must be zero. } (without loss of generality, we take $\kappa>0$). Substituting this solution into the BAEs~\eqref{BAEs}, 
the left hand side (LHS) of first equation in~\eqref{BAEs} is
\begin{equation}
    \frac{Z(i\kappa)}{e^{2(L+1)\kappa }}. 
\end{equation}
In the thermodynamic limit, the denominator of the above equation tends to infinity, 
while the corresponding right hand side (RHS) of first equation in~\eqref{BAEs} may take a finite value,
which leads to that $i\kappa$ is a pole of $Z(k)$, namely
\begin{equation}\label{cubic equation}
    \xi^3 + (\varepsilon_1 + \varepsilon_2)\xi^2 - \left( |V|^2 - \varepsilon_1\varepsilon_2 - 1 \right) \xi + \varepsilon_2 =0,
\end{equation}
with $\xi\equiv e^\kappa>1$. Numerical analysis shows that \eq\eqref{cubic equation} has always only one real root, which means that for a given set of model parameters, $\xi(p,U,\phi)$ (and $k_N$) is uniquely determined. The contribution of the imaginary Bethe root $k_N=i\kappa$ to the energy is given by $E_1=-\xi-\frac1\xi+\mu-\frac h2<-2+\mu-\frac h2$, which is lower than the energy of a real Bethe root. 
Thus, the condition $\xi(p,U,\phi)=1$ defines the phase boundary between the state with all real Bethe roots and the one-particle bound state. 

The spin rapidities $\lambda_\alpha$s of the ground state can be either entirely real, or one of the $\lambda_\alpha$s takes an imaginary value, corresponding to a spinon excitation in the spin sector. 
Substituting $k_N=i\kappa$ into the BAEs~\eqref{BAEs}, the other Bethe roots satisfy
\begin{subequations}\label{BEAs_BBS1}
 \begin{align}
    &e^{i k_j 2(L+1)} Z(k_j) 
      = \prod_{\alpha=1}^M e_{2u}(s_j + \lambda_\alpha) e_{2u}(s_j - \lambda_\alpha),
      \quad j=1,...,N-1, \label{BAEs_BBS1a}\\
    &e_{2u-2t}(\lambda_\alpha) e_{2u+2t}(\lambda_\alpha) \prod_{j=1}^{N-1} 
      e_{2u}(\lambda_\alpha - s_j) e_{2u}(\lambda_\alpha + s_j) \notag\\
    &\hspace{6.5em} = \prod_{\beta\neq\alpha}^M
      e_{4u}(\lambda_\alpha - \lambda_\beta) e_{4u}(\lambda_\alpha + \lambda_\beta),
      \quad \alpha=1,...,M,\label{BAEs_BBS1b}
\end{align}   
\end{subequations}  
with $e_n(x)=\frac{x+in/2}{x-in/2}$, $s_j=\sin k_j$ and 
$t=-i\sin k_N=\frac12(\xi-\frac1\xi)$. 
Assuming that $\lambda_1,...,\lambda_{M-1}$ are real and $\lambda_M=i\Lambda$ (we take $ \Lambda > 0 $ without loss of generality). Then Eq.\eqref{BAEs_BBS1b} with $\alpha=M$ becomes
\begin{equation}\label{imaginary_lambda}
    e_{2u-2t}(i\Lambda) e_{2u+2t}(i\Lambda) \prod_{j=1}^{N-1} 
    \frac{s_j^2+(\Lambda+u)^2}{s_j^2+(\Lambda-u)^2}
    = \prod_{\beta\neq\alpha}^{M-1}
    \frac{\lambda_\beta^2+(\Lambda+2u)^2}
    {\lambda_\beta^2+(\Lambda-2u)^2}.
\end{equation}
Similarly, in the thermodynamic limit, self-consistency of \eq\eqref{imaginary_lambda} demands that the boundary term $e_{2u-2t}(i\Lambda)e_{2u+2t}(i\Lambda)$ must be zero, which gives $\Lambda=t-u$ or $\Lambda=-t-u$.
The condition $\Lambda>0$ then selects $\Lambda=t-u$. Substituting $t=\frac12(\xi-\frac1\xi)$ into $\Lambda=t-u>0$, we obtain $\xi>\xi_1\equiv u+\sqrt{u^2+1}$.
We can prove that the formation of a spin string further lowers the energy (see Appendix \ref{appendix_a}). Thus, with $\xi(p,U,\phi)>\xi_1$, the ground state is characterized by the bound state with a spin string. 

An additional purely imaginary solution can exist in the charge sector. 
Substituting the Bethe roots $k_N=i\kappa$ and $\lambda_M=i\Lambda$ into BAEs~\eqref{BAEs}, 
the equations other real Bethe roots satisfying are 
\begin{equation} \label{BAEs_BBS2}
    \begin{split}
        &e^{i k_j 2(L+1)} Z(k_j) 
        =  e_{4u-2t}(s_j) e_{2t}(s_j)
        \prod_{\alpha=1}^{M-1}
        e_{2u}(s_j + \lambda_\alpha) e_{2u}(s_j - \lambda_\alpha),
        \ \ j=1,...,N-1,\\
        &e_{2u-2t}(\lambda_\alpha)e_{2t-6u}(\lambda_\alpha)
        \prod_{j=1}^{N-1} 
        e_{2u}(\lambda_\alpha - s_j) e_{2u}(\lambda_\alpha + s_j)\\
        &\hspace{6.5em} = \prod_{\beta\neq\alpha}^{M-1}
        e_{4u}(\lambda_\alpha - \lambda_\beta) 
        e_{4u}(\lambda_\alpha + \lambda_\beta),\ \ \alpha=1,...,M-1.
    \end{split}
\end{equation}
Assuming $k_{N-1}=i\eta$ ($\eta>0$ without loss of generality). By analogous analysis, $e^\eta$ satisfies
\begin{equation}
    \sinh\eta + \Lambda + u = 0 
    \quad \text{or} \quad 
    \sinh\eta - \Lambda + u = 0.
\end{equation}
Then $\eta>0$ gives $\sinh\eta = \Lambda - u > 0$ and the restriction
\begin{equation}
    \xi > \xi_2 \equiv 2u + \sqrt{4u^2+1}.
\end{equation}
Moreover, compared with real Bethe roots, the imaginary Bethe root 
$k_{N-1} = i\eta$ contributes a lower energy. Thus, $\xi = \xi_2$ marks the phase boundary for the emergence of the two-particle bound state. 

Increasing $\xi$ further does not generate additional bound states, since substituting 
$k_N = i\kappa$, $k_{N-1} = i\eta$, and $\lambda_M = i\Lambda$ into the BAEs~\eqref{BAEs} yields no boundary terms, which is consistent with physical intuition. 
\begin{figure}[h]
    \centering
    \includegraphics[width=0.75\columnwidth]{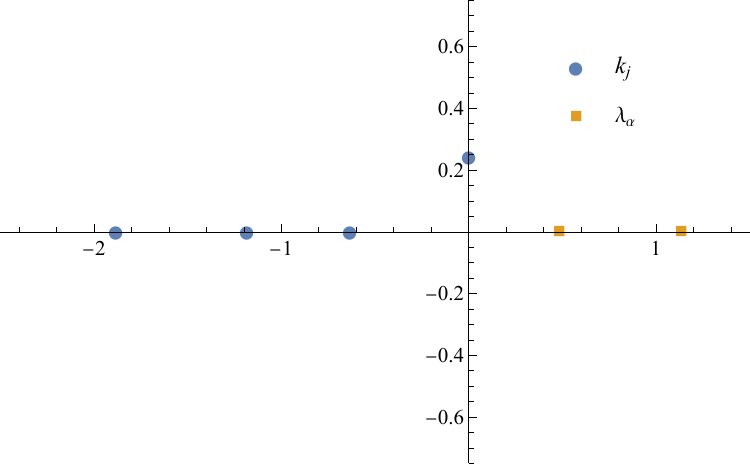}
    \caption{Numerical result of Bethe roots corresponding to the ground state where \(L=4\), \(N=4\), \(M=2\), \(p=1.3\), \(U=2\), \(\phi=0.5\). A purely imaginary Bethe root $k_N=i\kappa$ appears in charge sector.}
    \label{fig:bethe roots}
\end{figure}

So far, we have identified four distinct Bethe root configurations, which divide the ground state phase into four parameter regions, as shown below:
\begin{itemize}
    \item Region $\mathrm{I}$: $\xi < 1$, all Bethe roots are real and the ground state is free of boundary bound states.
    \item Region $\mathrm{II}$: $1 < \xi < \xi_1$, the one-particle boundary bound state appears with $k_N = i\kappa$ and all spin rapidities $\lambda_\alpha$ real. The corresponding ground state Bethe root pattern for a small-size system is shown in Fig.~\ref{fig:bethe roots}. 
    \item Region $\mathrm{III}$: $\xi_1 < \xi < \xi_2$, the ground state exhibits a boundary bound state in the charge sector ($k_N = i\kappa$), and a spin string forms ($\lambda_M = i\Lambda$).
    \item Region $\mathrm{IV}$: $\xi > \xi_2$, two-particle bound state forms with $k_N = i\kappa$, $k_{N-1} = i\eta$, and $\lambda_M = i\Lambda$. 
\end{itemize}
\begin{figure}[h]
    \centering
    \includegraphics[width=1\columnwidth]{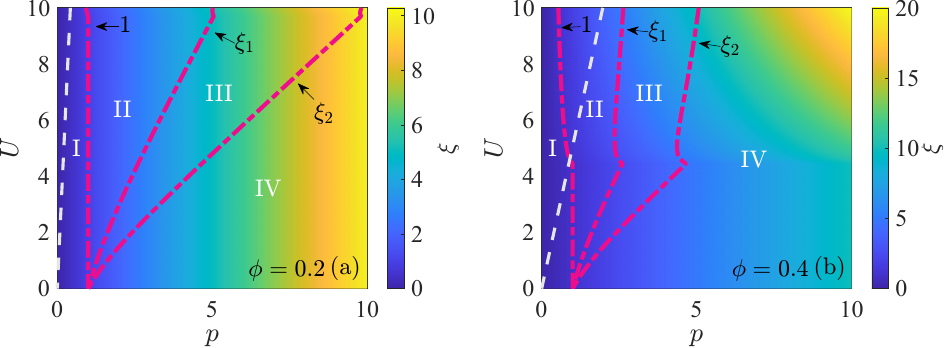}
    \caption{
Phase diagram of the ground state in the $(p,U)$ plane for (a) $\phi=0.2$ and (b) $\phi=0.4$, with both $p$ and $U$ ranging from 0 to 10. 
The color intensity represents the value of the parameter $\xi(p,U,\phi)$, which determines the ground state phase regions. 
Red dashed lines indicate the phase boundaries at the critical values $\xi = 1$, $\xi_1$, and $\xi_2$, separating the four distinct phases labeled $\mathrm{I}$–$\mathrm{IV}$ according to their Bethe root configurations. 
The white dashed line corresponds to $p = \frac{U}{4} \sinh^2(2\phi)$ and only the region to its right allows $\varepsilon_1,\varepsilon_2<0$. 
}
    \label{fig:parameter_space}
\end{figure}
Fig.~\ref{fig:parameter_space} shows the ground state phase diagram in the $(p,U)$ plane for fixed values of $\phi$. 
Increasing $p$ drives the ground state through regions II and III and ultimately into the two-particle bound state, corresponding to the gradual dominance of $|\varepsilon_1|$. Increasing $U$ drives the ground state toward regions I and II, corresponding to stronger localization of electrons. The parameter $\phi$ alters the phase boundaries, producing a more intricate phase structure. In particular, regions~II and~III correspond to configurations in which the impurity site is singly occupied, a necessary condition for the formation of a local magnetic moment. Determining whether a moment is actually realized, and the precise differences in physical properties between these two regions, requires further detailed analysis.

\section{Density of Bethe roots}
\label{sec:continuum limit}

In this section, we use the exact solution to perform a quantitative analysis of the impurity effects.
In the thermodynamic limit, the rapadities $k_j$ and $\lambda_\alpha$ tend to have continuous distribution with densities $\rho_c(k)$ and $\rho_s(k)$, respectively. Taking the logarithm of the BAEs and applying the Euler-Maclaurin formula \cite{Essler1996, Woynarovich1, Woynarovich2, Woynarovich3}, we obtain the following equations for the densities $\rho_c(k)$ and $\rho_s(k)$ corresponding to the ground state
\begin{align}
    \rho_c(k) &= \frac1\pi +\frac1L \hat{\rho}_c(k) 
    + \cos k \int_{-A}^A d\lambda \, \ a_{2u}(\sin k - \lambda) \rho_s(\lambda), \label{densities_int_eq1} \\
    \rho_s(\lambda) &= \frac{1}{L} \hat{\rho}_s(\lambda)
    + \int_{-Q}^Q dk\,a_{2u}(\lambda - \sin k) \rho_c(k) 
    - \int_{-A}^A d\lambda'\, a_{4u}(\lambda - \lambda') \rho_s(\lambda'), \label{densities_int_eq2}
\end{align}
where $a_n(x) = \frac{1}{2 \pi} \frac{n}{x^2+n^2/4}$ and the integration boundaries in the Euler-Maclaurin formula are replaced by $Q$ and $A$ with error of $o(L^{-2})$ \cite{Essler1996}. Higher order terms in the Euler-Maclaurin expansion are dropped. The driving terms of the $1/L$-corrections are given by
\begin{equation} 
    \hat{\rho}_c(k) = \frac{1}{\pi} -a_{2u}(\sin k) \cos k + \hat{\rho}_c^i(k)
\end{equation}
for the for the charge sector and
\begin{equation}
    \hat{\rho}_s(\lambda) = a_{4u}(\lambda) + \hat{\rho}_s^i(\lambda)
\end{equation}
for the spin sector, where
\begin{equation}\label{rho_c_hat_i}
    \hat{\rho}_c^i(k) = \frac{1}{2\pi i} \frac{d}{dk} \ln{Z(k)} +
\begin{cases}
    0 & \text{Regions } \mathrm{I} \text{ and }  \mathrm{II} \\ 
    \cos k \left[ a_{2t}(\sin k) + a_{4u-2t}(\sin k) \right] 
    & \text{Regions } \mathrm{III} \text{ and } \mathrm{IV}
\end{cases} 
\end{equation}
and
\begin{equation}
    \hat{\rho}_s^i (\lambda) = 
\begin{cases}
    0 & \text{Region } \mathrm{I} \\ 
    a_{2u+2t}(\lambda) + a_{2u-2t}(\lambda) & \text{Region } \mathrm{II} \\ 
    -a_{2t-2u}(\lambda) - a_{6u-2t}(\lambda) & \text{Region } \mathrm{III} \\ 
    0 & \text{Region } \mathrm{IV}
\end{cases} 
\end{equation}
are the densities induced by the impurity.
The boundaries $Q$ and $A$ are fixed by the conditions
\begin{equation}\label{boundary_fix_condition}
\begin{split}
    \int_{-Q}^Q dk \, \rho_c(k) 
    &= \frac{2 \left[ N - H(\xi-1) - H(\xi-\xi_2) \right] + 1}{L},\\
    \int_{-A}^A d\lambda \, \rho_s(\lambda) 
    &= \frac{2 \left[ M - H(\xi-\xi_1) \right] + 1}{L},\\
\end{split}
\end{equation}
where $H(x)$ is the Heaviside step function, defined as
\begin{equation}\label{Heaviside_function}
    H(x) = 
\begin{cases}
    0 & x \leq 0 \\ 
    1 & x > 0
\end{cases} \, .
\end{equation}

By virtue of the linearity of the coupled \eqs\eqref{densities_int_eq1} and~\eqref{densities_int_eq2}, the densities can be divided into the host part and the part of order of $1/L$.
\begin{equation}\label{rho_generic_solution}
    \rho = \rho^{\infty} + \frac{1}{L} (\rho^b + \rho^i).
\end{equation}
The first term in~\eqref{rho_generic_solution} is the host density in the limit $L\to\infty$, equivalent to the periodic boundary condition solution (see Ref.~\cite{Lieb1}). The remaining two terms, of order $1/L$, account for the impurity and boundary contributions, respectively. The terms $\rho^b$ and $\rho^i$ can be identified easily. In the absence of boundary potential and impurity (i.e., $p=0$ and $\varphi=0$), the $1/L$ terms reduce to $\rho^b$. The impurity component $\rho^i$ can be derived from $\rho$, $\rho^\infty$ and $\rho^b$. 
Using \eqs\eqref{boundary_fix_condition} and~\eqref{rho_generic_solution}, the electron density and the magnetization density can be expressed as follows:
\begin{equation}\label{electron_density}
    \frac{N}{L} = n^\infty + \frac1L (n^b + n^i) + o\left(\frac1L\right),
\end{equation}
with
\begin{equation}
\begin{split}
    &n^\infty = \frac12 \int_{-Q}^Q dk\, \rho_c^\infty(k), \quad
    n^b = \frac12 \int_{-Q}^Q dk\, \rho_c^b(k)-\frac12, \\
    &n^i = \frac12 \int_{-Q}^Q dk\, \rho_c^i(k) + H(\xi-1) + H(\xi-\xi_2)\label{n_i},
\end{split}
\end{equation}
and
\begin{equation}\label{magnetization_density}
    \frac{M^z}{L} = m^\infty + \frac1L (m^b + m^i)
    + o\left(\frac1L\right),
\end{equation}
with
\begin{equation}
\begin{split}
    &m^\infty = \frac14 \int_{-Q}^Q dk\, \rho_c^\infty(k) 
    - \frac12 \int_{-A}^A d\lambda\, \rho_s^\infty(\lambda),\quad
    m^b = \frac14 \int_{-Q}^Q dk\, \rho_c^b(k) 
    - \frac12 \int_{-A}^A d\lambda\, \rho_s^b(\lambda) +\frac14, \\
    &m^i = \frac14 \int_{-Q}^Q dk\, \rho_c^i(k)  
    - \frac12 \int_{-A}^A d\lambda\, \rho_s^i(\lambda) 
    + \frac12 H(\xi-1) - H(\xi-\xi_1) 
    + \frac12 H(\xi-\xi_2). \label{magnetization_density}
\end{split}
\end{equation}
Using \eqs\eqref{densities_int_eq1}, \eqref{densities_int_eq2}, and~\eqref{rho_generic_solution}, the densities induced by the impurity are given by
\begin{align}
    \rho_c^i(k) &=  \hat{\rho}_c^i(k) 
    + \cos k \int_{-A}^A d\lambda\, a_{2u}(\sin k - \lambda) \rho_s^i(\lambda), \label{rho_imp_int_eq1} \\
    \rho_s^i(\lambda) &= \hat{\rho}_s^i(\lambda)
    + \int_{-Q}^Q dk\, a_{2u}(\lambda - \sin k) \rho_c^i(k)
    - \int_{-A}^A d\lambda'\, a_{4u}(\lambda - \lambda') \rho_s^i(\lambda'). \label{rho_imp_int_eq2}
\end{align}

\section{Magnetic susceptibility in a weak field}
\label{sec:susceptibility}

In this section, we calculate the magnetization of the impurity and compare it with the host magnetization in a weak magnetic field.
The impurity contribution to the magnetization is given by \eq\eqref{magnetization_density}. From \eq\eqref{rho_imp_int_eq2}, the integral of $\rho_c^i$ is given in terms of $\rho_s^i$ as
\begin{equation}
    \int_{-Q}^Q dk\, \rho_c^i(k) = 2 \int_{-A}^Ad\lambda\, \rho_s^i(\lambda) + 2 \int_A^\infty d\lambda\, \rho_s^i(\lambda) - \left. \tilde{\hat{\rho}}_s^i(\omega) \right|_{\omega=0},
\end{equation}
with $\tilde{\hat{\rho}}_s^i(\omega)=\int_{-\infty}^\infty d\lambda\, e^{i\omega\lambda}\hat{\rho}_s^i(\lambda)$.
The magnetization density of the impurity in different parameter regions can now be written as
\begin{equation}\label{m_imp}
\begin{split}
    m^{i(\mathcal{R})} &= \frac12 \int_0^\infty d\lambda \, \rho_s^{i(\mathcal{R})} (\lambda + A) \\
    &= \frac12 \left. \tilde{g}^+(\omega) \right|_{\omega=0},
\end{split}
\end{equation}
where $\tilde{g}^+(\omega) = \int_0^\infty dz\, e^{i\omega z} g(z)$, $g(z)=\rho_s^{i(\mathcal{R})}(z+A)$, with index $\mathcal{R}$ labeling the ground state in regions I to IV. 
After Fourier-transforming, \eq\eqref{rho_imp_int_eq2} can be turned into a Wiener-Hopf equation \cite{Andrei2, Tsvelick1983, Frahm2006, Frahm1991correlation, Hubbard_book}
\begin{equation}
\begin{split}
    g(z) = &\ \hat{\rho}_{s,h=0}^{i(\mathcal{R})} (z+A) + \int_{-Q}^Q dk \, G_0(z+A-\sin k) \rho_c^{i(\mathcal{R})} (k)\\
    &+ \int_0^\infty dz' \, G_1(z-z') g(z') + \int_0^\infty dz' \, G_1(2A+z+z')g(z'),
\end{split}
\end{equation}
where
\begin{equation}\label{G_n}
    G_n(\lambda) = \frac{1}{2\pi} \int_{-\infty}^\infty d\omega \, 
    e^{-i\omega\lambda} \frac{e^{-nu |\omega| }}{2\cosh u\omega} ,
\end{equation}
and
\begin{equation}
    \hat{\rho}_{s,h=0}^{i(\mathcal{R})}(\lambda) = 
\begin{cases}
    0 & \text{Region } \mathrm{I} \\ 
    G_{\frac tu}(\lambda) + G_{-\frac tu}(\lambda) & \text{Region }\mathrm{II} \\ 
    - G_{\frac tu - 2}(\lambda) - G_{2-\frac tu}(\lambda) & \text{Region }\mathrm{III} \\ 
    0 & \text{Region }\mathrm{IV},
\end{cases}
\end{equation}
(see Appendix A). For a weak magnetic field, the above equation can be solved by iteration
$g(z)\simeq g_1(z)+g_2(z)$,
where
\begin{align}
    g_1(z) 
    &= \hat{\rho}_{s,h=0}^{i(\mathcal{R})} (z+A) 
    + C_i^{(\mathcal{R})} G_0(z+A) 
    + \int_0^\infty dz\, G_1(z-z')g_1(z')\label{g1},\\
    g_2(z) 
    &= \int_0^\infty dz'\, G_1(2A+z+z') g_1(z') 
    + \int_0^\infty dz'\, G_1(z-z') g_2(z')\label{g2}.
\end{align} 
The quantity $C_i^{(\mathcal{R})}$ is defined as $C_i^{(\mathcal{R})} \equiv \int_{-Q}^Q dk\, \exp \left( \frac{\pi\sin k}{2u} \right) \rho_c^{i(\mathcal{R})} (k)$. In \eq\eqref{g1}, we have used the fact that $A\gg1$ to approximate
\begin{equation}
    \int_{-Q}^Q dk \, G_0(z+A-\sin k) \rho_c^{i(\mathcal{R})} (k) \approx G_0(z+A) \int_{-Q}^Q dk\, 
    \exp \left( \frac{\pi\sin k}{2u} \right) 
    \rho_c^{i(\mathcal{R})} (k).
\end{equation}
In the weak-field limit, the density $\rho_c^{i(\mathcal{R})}(k)$ is obtained by analogy to Ref.~\cite{Asakawa2} as the solution to the integral equation (see Appendix B)
\begin{equation}\label{rho_c_imp_int_eq}
\begin{split}
    \rho_c^{i(\mathcal{R})} (k) 
    \simeq &\ \hat{\rho}_{c,h=0}^{i(\mathcal{R})} (k) 
    - \frac1u \cos k \cosh \left( \frac{\pi \sin k}{2u} \right) 
    \exp \left( -\frac{\pi A}{2u} \right) \tilde{g}^+ \left( \frac{i \pi}{2u} \right) \\
    &+ \cos k \int_{-Q}^Q dk'\, G_1(\sin k - \sin k') \rho_c^{i(\mathcal{R})} (k'),
\end{split}
\end{equation}
where $\hat{\rho}_{c,h=0}^{i(\mathcal{R})}(k)$ takes the values of 
\begin{equation}\label{hat_rho_c_R_h0}
    \hat{\rho}_{c,h=0}^{i(\mathcal{R})}(k) = \hat{\rho}_c^i(k) +
\begin{cases}
        0 & \text{Region } \mathrm{I}\\ 
    \cos k \left[ G_{\frac{t}{u}+1}(\sin k) + G_{1-\frac{t}{u}}(\sin k) \right]  & \text{Region } \mathrm{II} \\ 
    \cos k \left[ - G_{\frac{t}{u}-1}(\sin k) 
    - G_{3-\frac{t}{u}}(\sin k) \right]  & \text{Region } \mathrm{III}\\ 
    0 & \text{Region } \mathrm{IV}
\end{cases} \, ,
\end{equation}
and $\hat{\rho}_c^i(k)$ is given by \eq\eqref{rho_c_hat_i}.
\eqs\eqref{g1} and~\eqref{g2} can now be solved by means of Wiener-Hopf techniques. The results for $g_1$ and $g_2$ are obtained (see Appendix C) 
\begin{align}
    \tilde{g}_1^+(\omega) &\simeq \frac{i}{2u} 
    \left(C_i^{(\mathcal{R})} + 2 \Theta^{(\mathcal{R})} \cos\frac{\pi t}{2u} \right) 
    \exp\left(-\frac{\pi A}{2u}\right) 
    G^+(\omega)
    \frac{G^-(-\frac{i \pi}{2u})}{\omega + \frac{i \pi}{2u}} \label{g1_p}\\
    \tilde{g}_2^+(\omega) &\simeq \frac{u}{\pi \sqrt{\pi e}} \left(C_i^{(\mathcal{R})} + 2 \Theta^{(\mathcal{R})} 
    \cos\frac{\pi t}{2u} \right)
    \exp\left(-\frac{\pi A}{2u}\right) G^+(\omega)
    \int_0^\infty dx\, \frac{e^{-2Ax}}{x - i\omega} x, \label{g2_p}
\end{align}
where
\begin{equation}\label{Theta_function}
    \Theta^{(\mathcal{R})} = 
\begin{cases}
    0 & \text{Regions } \mathrm{I} \text{ and }  \mathrm{IV} \\
    1 & \text{Regions } \mathrm{II} \text{ and } \mathrm{III}
\end{cases} \, ,
\end{equation}
and the functions $G^{\pm}(\omega)$ take the form \cite{Frahm1991correlation, Hubbard_book}
\begin{equation}\label{G_pm}
    G^+(\omega) = G^-(-\omega) 
    = \frac{\sqrt{2 \pi}}{\Gamma(\frac12 - i\frac{u \omega}{\pi})}
    \left( -i\frac{u \omega}{\pi e} \right)^{-i u \omega/\pi},
\end{equation}
with $\Gamma(\omega)$ denoting the gamma function.
Using \eqs\eqref{m_imp}, \eqref{g1_p}, and~\eqref{g2_p}, the leading behavior of the impurity magnetization is given by
\begin{equation}\label{impurity_incuced_magnetization}
    m^i \simeq 
\begin{cases}
    \frac{1}{\sqrt{2 \pi e}} C_i^{(\mathrm{I})} \exp\left( -\frac{\pi A}{2u} \right) \left( 1 + \frac{u}{2 \pi A} \right) & \text{Region }\mathrm{I} \\
    \frac{1}{\sqrt{2 \pi e}} \left( C_i^{(\mathrm{II})} + 2\cos{\frac{\pi t}{2u}} \right) \exp\left( -\frac{\pi A}{2u} \right) \left( 1 + \frac{u}{2 \pi A} \right) & \text{Region } \mathrm{II}\\
    \frac{1}{\sqrt{2 \pi e}} \left( C_i^{(\mathrm{III})} + 2\cos{\frac{\pi t}{2u}} \right) \exp\left( -\frac{\pi A}{2u} \right) \left( 1 + \frac{u}{2 \pi A} \right) & \text{Region }\mathrm{III} \\
    \frac{1}{\sqrt{2 \pi e}} C_i^{(\mathrm{IV})} \exp\left( -\frac{\pi A}{2u} \right) \left( 1 + \frac{u}{2 \pi A} \right) & \text{Region }\mathrm{IV}
\end{cases} \, .
\end{equation}

To calculate the impurity's magnetic susceptibility, the boundaries $A$ must be determined as a function of the magnetic field through the condition $\varepsilon_s(A)=0$ on the dressed energy. 
Here $\varepsilon_s(\lambda)$ is the solution of the coupled integral equations \cite{Frahm1990critical, Frahm1991correlation}
\begin{align}
    \varepsilon_c(k) &= \varepsilon_c^{(0)}(k) 
    + \int_{-A}^A d\lambda'\, a_{2u}(\sin k - \lambda') \varepsilon_s(\lambda'), \label{dressed_energy1} \\
    \varepsilon_s(\lambda) &= \varepsilon_s^{(0)}(\lambda)
    + \int_{-Q}^Q dk\, \cos k \,a_{2u}(\lambda - \sin k) \varepsilon_c(k) 
    - \int_{-A}^A d\lambda'\, a_{4u}(\lambda - \lambda') \varepsilon_s(\lambda'), \label{dressed_energy2}
\end{align}
where
\begin{equation}
    \varepsilon_c^{(0)}(k) = \mu - \frac{h}{2} - 2 \cos k, \quad
    \varepsilon_s^{(0)}(\lambda) = h.
\end{equation}
In the ground state, the boundary $Q$ is determined by the chemical potential and the magnetic field via the condition $\varepsilon_c(Q)=0$. A Wiener-Hopf analysis gives
\begin{equation}\label{pi_A}
   \pi A \simeq -2u\ln(h/h_0) + \frac{u}{2} \frac{1}{\ln(h/h_0)},
\end{equation}
where $h_0 = -\frac{C}{2u} \sqrt{\frac{2 \pi}{e}}$. The quantity $C$ is defined as
\begin{equation}
    C=\int_{-Q}^Q dk\, \cos k \exp\left(\frac{\pi \sin k}{2u}\right) \varepsilon_c(k).
\end{equation}
Following Ref.~\cite{Asakawa2}, under the condition that $A \gg 1$, the dressed energy $\varepsilon_c(k)$ in \eqs\eqref{dressed_energy1} and~\eqref{dressed_energy2} is determined as the solution to the following integral equation:
\begin{equation}\label{dressed_energy_C}
\begin{split}
    \varepsilon_c(k) \simeq &\ \mu - 2 \cos k 
    - \frac1u \cosh\left( \frac{\pi \sin k}{2u} \right) \exp\left( -\frac{\pi A}{2u} \right) 
    \tilde{y}^+ \left( \frac{i \pi}{2u} \right) \\
    &+ \int_{-Q}^Q dk'\, G_1(\sin k -\sin k') \cos k' \varepsilon_c(k'), 
\end{split}
\end{equation}
where $\tilde{y}^+(\omega)\simeq \tilde{y}_1^+(\omega)+\tilde{y}_2^+(\omega)$, and
\begin{align}
    \tilde{y}_1^+(\omega) &\simeq \frac{i}{2} G^+(\omega) 
    \left[ 
    \frac{hG^-(0)}{\omega + i0} 
    + C \frac1u \frac{G^-(-\frac{i\pi}{2u})}{\omega + \frac{i\pi}{2u}} 
    \exp \left( -\frac{\pi A}{2u} \right)
    \right], \label{y1}\\
    \tilde{y}_2^+(\omega) &\simeq \frac{uh}{\sqrt{2}\pi} G^+(\omega) 
    \int_0^\infty dx\, \frac{e^{-2Ax}}{x - i \omega}. \label{y2}
\end{align}
To compare the susceptibility induced by the impurity with the bulk susceptibility, the bulk magnetization $m^\infty$ in a weak magnetic field is given by the same procedure used in deriving $m^i$
\begin{equation}\label{bulk_magnetization}
    m^\infty \simeq \frac{1}{\sqrt{2 \pi e}}
    \exp\left( -\frac{\pi A}{2u} \right) 
    \left( 1 + \frac{u}{2 \pi A} \right)
    \int_{-Q}^Q dk\, \exp\left( \frac{\pi \sin k}{2u} \right) \rho_c^\infty(k),
\end{equation}
where the density $\rho_c^\infty(k)$ is given in terms of the integral equation
\begin{equation}\label{bulk_density_of_Bethe_roots}
\begin{split}
    \rho_c^\infty(k) \simeq \, &\frac1\pi 
    + \cos k \int_{-Q}^Q dk'\, 
    G_1(\sin k - \sin k') \rho_c^\infty(k')\\
    &- \frac{1}{2eu} \cos k \cosh\left( \frac{\pi \sin k}{2u} \right)
    \exp \left( -\frac{\pi A}{u} \right) 
    \int_{-Q}^Q dk'\, \exp\left( \frac{\pi\sin k'}{2u} \right) \rho_c^\infty(k') .
\end{split}
\end{equation}
\begin{figure}[h]
    \centering
    \includegraphics[width=0.6\linewidth]{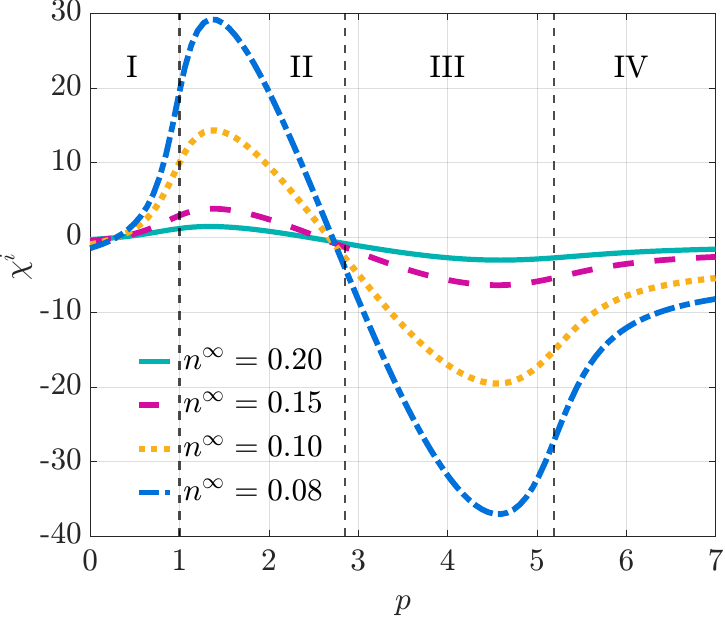}
    \caption{Ground state impurity susceptibility $\chi^i$ at low electron density, shown as a function of $p$ for $U=5$ and $\phi=0.3$.  
Each curve corresponds to a different electron density $n^\infty$. 
The labels $\mathrm{I}$–$\mathrm{IV}$ indicate the four distinct phase regions, separated by black dashed lines. 
The maximum of $\chi^i$ occurs in region $\mathrm{II}$. }
    \label{fig:chi_vs_p}
\end{figure}

\begin{figure}
    \centering
    \includegraphics[width=0.5\linewidth]{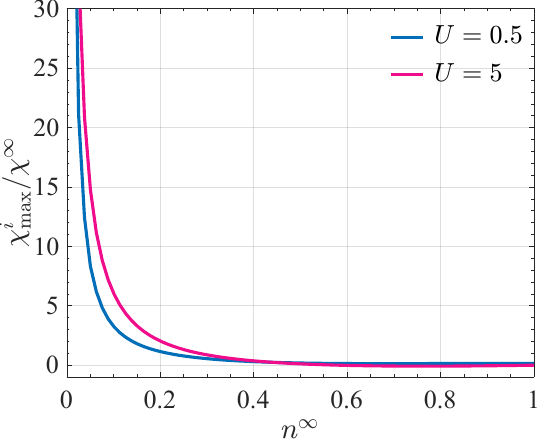}
    \caption{The ratio $\chi^i_{\max}/\chi^\infty$ as a function of the particle density $n^\infty$ for $\phi=0.3$.}
    \label{fig:ratio_chi}
\end{figure}

Using \eqs\eqref{impurity_incuced_magnetization}, \eqref{pi_A}, and~\eqref{bulk_magnetization}, we compute the impurity-induced susceptibility $\chi^i=\frac{\partial m^i}{\partial h}$ and the host susceptibility $\chi^\infty=\frac{\partial m^\infty}{\partial h}$. 
As shown in Fig.~\ref{fig:chi_vs_p}, $\chi^i$ exhibits a peak in region II ( denoted $\chi^i_{\max}$) which grows as the electron density $n^\infty$ decreases. 
Fig.~\ref{fig:ratio_chi} further shows that the ratio $\chi^i_{\max}/\chi^\infty$ diverges in the dilute limit $n^\infty \to 0$ following the Curie law $\chi \propto 1/T$, while it approaches a finite value as $n^\infty \to 1$.  
Consistently, \eq\eqref{n_i} indicates that in region II the impurity-induced electron number $n^i$ approaches one as $n^\infty \to 0$. 
These results consistently reflect the formation of a localized moment at the impurity site. 
At high electron densities, the impurity spin is effectively screened by the conduction electrons, consistent with Kondo physics \cite{Kondo1970,Andrei2},  
whereas in the dilute limit the screening is strongly suppressed, leading to the localization of a single electron at the impurity site. 




\section{Conclusion}
\label{conclusion}
In this work, we have constructed and analyzed a strongly correlated one-dimensional electron model with a single Anderson impurity embedded at the boundary. By solving the model exactly, we determine the ground state phase diagram, identifying impurity-induced boundary bound states across different parameter regimes. Our analysis shows that in the dilute electron limit, a localized magnetic moment forms at the impurity site, as indicated by the divergence of the impurity-induced susceptibility, whereas at finite electron densities, the moment is partially screened by the host electrons, consistent with Kondo physics. These results provide a rigorous and exact characterization of impurity-induced magnetic properties in a strongly correlated host, serving as a benchmark for future studies of boundary impurities in many-body systems.

\section*{Acknowledgment}
\addcontentsline{toc}{section}{Acknowledgment}
Renjie Song acknowledges Pei Sun for valuable advice and Xin Zhang for helpful discussions. We acknowledge the financial supports from the National Key R\&D Program of China (Grant No. 2021YFA1402104), National Natural Science Foundation of China (Grant No. 12447130, 12205235, 12105221, 12434006, 12247103, 12247179 and 12175180), and Postdoctoral Fellowship Program of CPSF (Grant No. GZC20252261).

\begin{appendix}

\section{Impurity effects}
\label{appendix_a}
\begin{figure}
    \centering
    \includegraphics[width=0.6\linewidth]{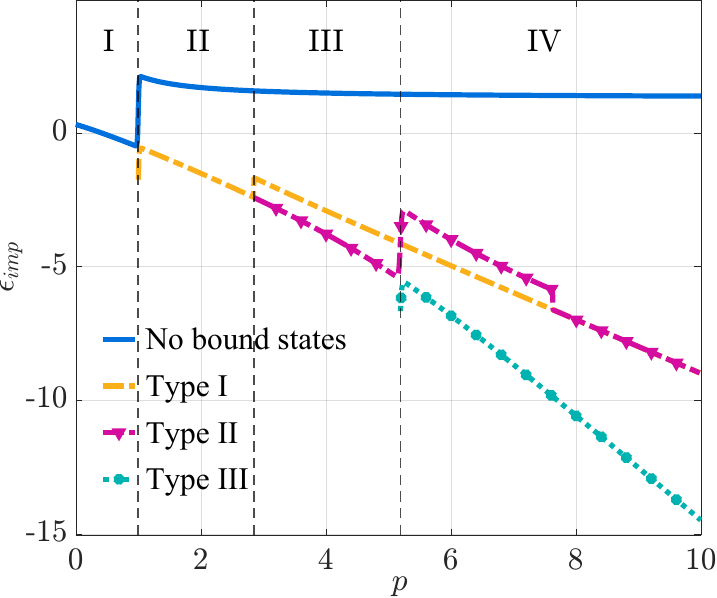}
    \caption{Impurity contribution to the ground state energy 
    for various state configurations as a function of the parameter $p$ for fixed $U=5$, $\phi=0.3$ and electron density $n^\infty=0.7$. Regions $\mathrm{I}$–$\mathrm{IV}$ correspond to four phases, with boundaries marked by the black dashed lines $\xi(p,U,\phi)=1$, $\xi_1$, and $\xi_2$.}
    \label{fig:impurity_energy}
\end{figure}

We have identified four parameter regions, three of which correspond to distinct bound state configurations.
For convenience, we classify the impurity bound states into three types: Type I, corresponding to the one-particle bound state; Type II, corresponding to the one-particle bound state with a spin string; and Type III, corresponding to the two-particle bound state.
In this appendix, we give a numerical calculation of the impurity's energy contribution. By using the decomposition \eqref{rho_generic_solution}, the host, impurity and boundary contributions to any thermodynamic quantity can be treated separately. Consequently, the energy per site is given by
\begin{equation}
    \frac{E_{g}}{L} = \epsilon_0 + \frac{\epsilon_{imp} + \epsilon_b}{L}.
\end{equation}
The expression for the impurity contribution in terms of the densities is
\begin{equation}\label{impurity_energy}
    \epsilon_{imp} 
    = \frac12 \int_{-Q}^Q dk\, (\mu - \frac h2 - 2 \cos k)\rho_c^i(k) +
\begin{cases}
        0 & \text{No bound states} \\ 
    E_1 & \text{Type } \mathrm{I}\\ 
    E_1 + h & \text{Type } \mathrm{II}\\ 
    E_1 + E_2 + h& \text{Type } \mathrm{III}
\end{cases} \,,
\end{equation}
where $E_1=-\xi-\frac1\xi+\mu-\frac h2$ and $E_2=-2\sqrt{1+(t-2u)^2}+\mu-\frac h2$. 
In the absence of a magnetic field, \eqs\eqref{rho_imp_int_eq1} and~\eqref{rho_imp_int_eq2} can be rewritten via Fourier transformation as follows:
\begin{align}
    \rho_{c,h=0}^i(k) &= \hat{\rho}_{c,h=0}^i(k) 
    + \cos k \int_{-Q}^Q dk' \, G_1(\sin k - \sin k') \rho_{c,h=0}^i (k'), \label{rho_imp_int_eq1_h0} \\
    \rho_{s,h=0}^i(\lambda) &= \hat{\rho}_{s,h=0}^i(\lambda) 
    + \int_{-Q}^Q dk \, G_0(\lambda - \sin k) \rho_{c,h=0}^i (k).\label{rho_imp_int_eq12_h0}
\end{align}
The densities are determined by the solution to a scalar integral equation for the charge rapidity density $\rho_{c,h=0}^i(k)$. 
The driving terms in different regions are given by
\begin{equation}\label{rho_c_hat_h0}
    \hat{\rho}_{c,h=0}^i(k) = \hat{\rho}_c^i(k) +
\begin{cases}
        0 & \text{No bound states} \\ 
    \cos k \left[ G_{\frac{t}{u}+1}(\sin k) + F^{(1)}_{\frac{t}{u}}(\sin k) \right]  & \text{Type } \mathrm{I} \\ 
    \cos k \left[ - G_{\frac{t}{u}-1}(\sin k) 
    - F^{(1)}_{\frac{t}{u}-2}(\sin k) \right]  & \text{Type } \mathrm{II}\\ 
    0 & \text{Type } \mathrm{III}
\end{cases} \,,
\end{equation}
with $\hat{\rho}_c^i$ given by \eq\eqref{rho_c_hat_i}, and
\begin{equation}
    \hat{\rho}_{s,h=0}^i(\lambda) = 
\begin{cases}
    0 & \text{No bound states} \\ 
    G_{\frac tu}(\lambda) + F^{(0)}_{\frac tu}(\lambda) & \text{Type } \mathrm{I} \\ 
    - G_{\frac tu - 2}(\lambda) - F^{(0)}_{\frac tu -2}(\lambda) & \text{Type } \mathrm{II} \\ 
    0 & \text{Type } \mathrm{III}
\end{cases} \, .
\end{equation}
Here, the function $F^{(m)}_n(\lambda)$ is defined as
\begin{equation}
    F^{(m)}_n(\lambda) = 
\begin{cases}
    G_{-n+m}(\lambda) & n<1 \\ 
    0 & n=1 \\ 
    - G_{n-2+m}(\lambda) & n>1
\end{cases}\, .
\end{equation}
The ground state configuration is that which minimizes $\epsilon_{imp}$. Fig.~\ref{fig:impurity_energy} illustrates that for $\xi(p,U,\phi)>1$ (regions $\mathrm{II}$–$\mathrm{IV}$), corresponding to an attractive boundary potential sufficiently strong to induce binding, the ground state is the bound state that first emerges in the respective parameter region.

\section{Thermodynamic limit} 
\label{appendix_c}
Following Ref.~\cite{Asakawa2}, a detailed derivation of \eq\eqref{rho_c_imp_int_eq} is given as follows.
By Fourier transforming, \eq\eqref{rho_imp_int_eq2} gives
\begin{equation}\label{tilde_rho_s}
\begin{split}
    \tilde{\rho}_s^i(\omega) 
    =&\ \frac{1}{1 + e^{-2u|\omega|}} \tilde{\hat{\rho}}_s^i(\omega)
    + \frac{e^{-u|\omega|}}{1 + e^{-2u|\omega|}}
    \int_{-Q}^Q dk\, e^{i\omega\sin k} \rho_c^i(k) \\
    &+ \frac{e^{-2u|\omega|}}{1 + e^{-2u|\omega|}} 
    \int_{|\lambda|>A} d\lambda\, e^{i\omega\lambda} \rho_s^i(\lambda).
\end{split}
\end{equation}
Using \eq\eqref{tilde_rho_s}, the integral in \eq\eqref{rho_imp_int_eq1} can be rewritten as
\begin{align}\label{integral_a2u_1}
    &\int_{-A}^A d\lambda\, a_{2u}(\sin k - \lambda) \rho_s^i(\lambda) \notag\\ 
    = &\int_{-\infty}^\infty d\lambda\,
    a_{2u}(\sin k - \lambda) \rho_s^i(\lambda)
    - \int_{|\lambda|>A} d\lambda\, 
    a_{2u}(\sin k - \lambda) \rho_s^i(\lambda) \notag \\ 
    = &\frac{1}{2\pi} \int_{-\infty}^\infty d\omega\,
    e^{i\omega\sin k} \tilde{a}_{2u}(\omega) \tilde{\rho}_s^i(\omega)
    - \int_{|\lambda|>A} d\lambda\, 
    a_{2u}(\sin k - \lambda) \rho_s^i(\lambda) \notag \\ 
    = &\frac{1}{2\pi} \int_{-\infty}^\infty d\omega\,
    \frac{e^{-i\omega \sin k}}{2 \cosh u\omega}
    \tilde{\hat{\rho}}_s^i(\omega) 
    + \int_{-Q}^Q dk'\, G_1(\sin k - \sin k') \rho_c^i(k') \notag\\ 
    &+ \int_{|\lambda|>A} d\lambda\,
    G_2(\sin k - \lambda) \rho_s^i(\lambda)
    - \int_{|\lambda|>A} d\lambda\, 
    a_{2u}(\sin k - \lambda) \rho_s^i(\lambda) \notag \\
    = &\int_{-\infty}^\infty d\lambda\, 
    G_0(\sin k - \lambda) \hat{\rho}_s^i(\lambda) 
    + \int_{-Q}^Q dk'\, G_1(\sin k - \sin k') \rho_c^i(k') \notag\\ 
    &+ \int_{|\lambda|>A} d\lambda\,
    G_2(\sin k - \lambda) \rho_s^i(\lambda)
    - \int_{|\lambda|>A} d\lambda\, 
    a_{2u}(\sin k - \lambda) \rho_s^i(\lambda) .
\end{align}
Here, we have used the fact that $\tilde{\rho}_s^i(\omega)=\tilde{\rho}_s^i(-\omega)$, $G_n^*(\lambda)=G_n(-\lambda)=G_n(\lambda)$, and Parseval's identity
\begin{equation}
    \int_{-\infty}^\infty d\lambda\, \mathcal{F}(\lambda) \mathcal{G}^*(\lambda)
    = \frac{1}{2\pi} \int_{-\infty}^\infty d\omega\, 
    \tilde{\mathcal{F}}(\omega) \tilde{\mathcal{G}}^*(\omega),
\end{equation}
where $\mathcal{G}^*$ denotes the complex conjugate of $\mathcal{G}$.
The Fourier transform gives
\begin{equation}
    \tilde{G}_2(\omega) - \tilde{a}_{2u}(\omega) = -\tilde{G}_0(\omega),
\end{equation}
which implies
\begin{equation}\label{G2_relation}
    G_2(\lambda) - a_{2u}(\lambda) = -G_0(\lambda).
\end{equation}
Substituting \eq\eqref{G2_relation} into \eq\eqref{integral_a2u_1}, we get
\begin{equation}\label{integral_a2u_2}
\begin{split}
    \int_{-A}^A d\lambda\, a_{2u}(\sin k - \lambda) \rho_s^i\lambda)
    = &\int_{-\infty}^\infty d\lambda\,
    G_0(\sin k - \lambda) \hat{\rho}_s^i(\lambda)
    + \int_{-Q}^Q dk'\, G_1(\sin k - \sin k') \rho_c^i(k') \\
    &- \int_{|\lambda|>A} d\lambda\,
    G_0(\sin k - \lambda) \rho_s^i(\lambda).
\end{split}
\end{equation}
Note that
\begin{equation}
    G_0(\lambda)=\frac{1}{4u \cosh \frac{\pi \lambda}{2u}},
\end{equation}
and the last term in \eq\eqref{integral_a2u_2} can therefore be approximated in the weak-field limit ($A\gg1$)
\begin{equation}
\begin{split}
    &\int_{|\lambda|>A} d\lambda\,
    G_0(\sin k - \lambda) \rho_s^i(\lambda)\\
    = &\frac{1}{4u} \int_0^\infty d\lambda\,
    \left[ 
    \frac{1}{\cosh \frac{\pi}{2u} (A + \lambda - \sin k)}
    + \frac{1}{\cosh \frac{\pi}{2u} (A + \lambda + \sin k)}
    \right] 
    \rho_s^i(\lambda + A)\\
    \simeq & \frac{1}{u} 
    \cosh \left( \frac{\pi \sin k}{2u} \right)
    \exp \left( -\frac{\pi A}{2u} \right)
    \int_0^\infty d\lambda\,
    \exp \left( -\frac{\pi \lambda}{2u} \right) g(\lambda)\\
    = &\frac 1u 
    \cosh \left( \frac{\pi \sin k}{2u} \right)
    \exp \left( -\frac{\pi A}{2u} \right) 
    \tilde{g}^+ \left( \frac{i\pi}{2u} \right), 
\end{split}
\end{equation}
where $\tilde{g}^+(\omega)\simeq \tilde{g}_1^+(\omega)+\tilde{g}_2^+(\omega)$, $\tilde{g}_1^+(\omega)$ and $\tilde{g}_2^+(\omega)$ are given by \eqs\eqref{g1_p} and~\eqref{g2_p}, respectively.
Thus \eq\eqref{integral_a2u_2} can now be rewritten as
\begin{equation}\label{integral_a2u_3}
 \begin{split}
    \int_{-A}^A d\lambda\, a_{2u}(\sin k - \lambda) \rho_s^i(\lambda)
    \simeq &\int_{-\infty}^\infty d\lambda\,
    G_0(\sin k - \lambda) \hat{\rho}_s^i(\lambda)
    + \int_{-Q}^Q dk'\, G_1(\sin k - \sin k') \rho_c^i(k') \\
    &- \frac 1u 
    \cosh \left( \frac{\pi \sin k}{2u} \right)
    \exp \left( -\frac{\pi A}{2u} \right) 
    \tilde{g}^+ \left( \frac{i\pi}{2u} \right) .
\end{split}   
\end{equation}
Using \eq\eqref{integral_a2u_3}, the integral equation \eqref{rho_imp_int_eq1} in region $\mathcal{R}$ can be decoupled as 
\begin{equation}
\begin{split}
    \rho_c^i(k) 
    \simeq &\ \hat{\rho}_c^i(k) 
    - \frac1u \cos k \cosh \left( \frac{\pi \sin k}{2u} \right) 
    \exp \left( -\frac{\pi A}{2u} \right) \tilde{g}^+ \left( \frac{i \pi}{2u} \right) \\
    &+ \cos k \int_{-\infty}^\infty d\lambda\,
    G_0(\sin k - \lambda) \hat{\rho}_s^i(\lambda) \\
    &+ \cos k \int_{-Q}^Q dk'\, G_1(\sin k - \sin k') \rho_c^i(k') \\
    = &\ \hat{\rho}_{c,h=0}^i(k) 
    - \frac1u \cos k \cosh \left( \frac{\pi \sin k}{2u} \right) 
    \exp \left( -\frac{\pi A}{2u} \right) \tilde{g}^+ \left( \frac{i \pi}{2u} \right) \\
    &+ \cos k \int_{-Q}^Q dk'\, G_1(\sin k - \sin k') \rho_c^i(k'),
\end{split}
\end{equation}
with $\hat{\rho}_{c,h=0}^i(k)$ given by  \eq\eqref{hat_rho_c_R_h0}.

\section{Wiener-Hopf equation}
\label{appendix_b}
The integral \eqs\eqref{g1} and \eqref{g2} are of Wiener-Hopf type
\begin{equation}\label{WH_equaton}
    g(z) = g^{(0)}(z) + \int_0^\infty dz'\, G_1(z-z') g(z').
\end{equation}
After Fourier transforming and factorizing the kernel as
\begin{equation}
\left[ 1 - G_1(\omega) \right]^{-1} = G^+(\omega) G^-(\omega), \quad
\lim_{\omega\to\infty} G^{\pm}(\omega) = 1,
\end{equation}
with $G^{\pm}(\omega)$ analytic for $\pm\operatorname{Im}(\omega) > 0$,
\eq\eqref{WH_equaton} becomes
\begin{equation}\label{WH_equation_factorization_1}
\left[ G^+(\omega) \right]^{-1} \tilde{g}^+(\omega) + G^-(\omega) \tilde{g}^-(\omega)
= G^-(\omega) \tilde{g}^{(0)}(\omega),
\end{equation}
where $\tilde{g}^{\pm} (\omega) = \int_{-\infty}^\infty dz\,e^{i\omega z} H(\pm z)g(z)$ and $H(z)$ is defined as  \eq\eqref{Heaviside_function}. Decomposing $G^-(\omega)\tilde{g}^{(0)}(\omega)$ into $Q^{\pm}(\omega)$ analytic in the upper and lower half-planes yields
\begin{equation}\label{WH_equation_factorization_2}
\tilde{g}^+(\omega) = G^+(\omega) Q^+(\omega), \quad
\tilde{g}^-(\omega) = \frac{Q^-(\omega)}{G^-(\omega)}.
\end{equation}
For the Hubbard model, the kernel factorization is given in \eq\eqref{G_pm}.

This paper considers three forms of the driving term $g_\alpha^{(0)}(z)$, with $\alpha$ labeling the different cases.

(a) $g_a^{(0)}(z)=G_0(z+A)$.
The Fourier transform of the driving term has simple poles at $\omega_n=i(2n+1)\pi/2u$, $n\in\mathbb{Z}$.
To determine the decomposition into $Q_a^+(\omega)$, the residues of 
$G^-(\omega)\tilde{g}_a^{(0)}(\omega)$ that correspond to the part analytic in the upper half-plane are given by
\begin{equation}
\begin{split}
    &\lim_{\substack{\omega\to -\omega_n \\ (n \geq 0)}}
    \left[ (\omega + \omega_n) G^-(\omega) \tilde{g}_a^{(0)}(\omega) \right] \\
    &= \frac{i}{2 \pi} (-1)^n G^-\left[ -i\frac{\pi}{2u} (2n+1) \right] 
    \exp\left[ {-\frac{\pi}{2u} (2n+1) A} \right] \quad (n\geq0).
\end{split}
\end{equation}
Then $Q_a^+(\omega)$ can be determined
\begin{equation}
    Q_a^+(\omega) = \frac{i}{2u} \sum_{n=0}^\infty (-1)^n 
    \frac{G^-\left[ -i \frac{\pi}{2u} (2n+1) \right]}{\omega + i \frac{\pi}{2u} (2n+1)}
    \exp \left[-\frac{\pi}{2u} (2n+1) A \right].
\end{equation}
From \eq\eqref{WH_equation_factorization_2}, the leading term of $\tilde{g}_a^+(\omega)$ reads
\begin{equation}\label{g_case_a}
    \tilde{g}_a^+(\omega) \simeq \frac{i}{2u} G^+(\omega)
    \frac{ G^-\left( -i \frac{\pi}{2u} \right) }{ \omega + i \frac{\pi}{2u} }
    \exp\left( -\frac{\pi A}{2u} \right).
\end{equation}

(b) $g_b^{(0)}(z)=G_m(z+A)+G_{-m}(z+A)$.
Following a procedure analogous to that in the first case, the leading term of $\tilde{g}_b^+(\omega)$ takes the form
\begin{equation}\label{g_case_b}
    \tilde{g}_b^+(\omega) \simeq \frac{i}{u} G^+(\omega)
    \frac{ G^-\left( -i \frac{\pi}{2u} \right) }{ \omega + i \frac{\pi}{2u} }
    \exp\left( -\frac{\pi A}{2u} \right) \cos \frac{m\pi}{2}.
\end{equation}

(c) $g_c^{(0)}(z)=\int_0^\infty dz'\, G_1(2A+z+z')g_1(z')$.
The Fourier transform of $g_c^{(0)}(z)$ is 
\begin{equation}
    \tilde{g}_c^{(0)}(\omega) = \left[ 1 - \frac{1}{G^+(\omega)G^-(\omega)} \right] 
    e^{-i2 \omega A} \tilde{g}_1^+(-\omega),
\end{equation} 
where $\tilde{g}_1^+(\omega)$ is the component of the Fourier transform of $g_1(z)$ analytic in the upper half-plane, with $g_1(z)$ defined in \eq\eqref{g1}. 
Using \eqs\eqref{g_case_a} and~\eqref{g_case_b}, the leading term is given by
\begin{equation}
\begin{split}
    \tilde{g}_1^+(\omega) \simeq 
    \frac{i}{2u} G^+(\omega) 
    \frac{ G^-\left(-i \frac{\pi}{2u}\right) }{ \omega + i \frac{\pi}{2u} }
    \exp\left(-\frac{\pi A}{2u}\right) 
    \left(
    C_i^{(\mathcal{R})} 
    + 2 \Theta^{(\mathcal{R})}\cos \frac{t \pi}{2u} 
    \right).
\end{split}
\end{equation}
Following \cite{Hubbard_book}, the function $Q_c^+(\omega)$ can be expressed as
\begin{equation}\label{Q_2_plus}
\begin{split}
    Q_c^+(\omega) = &\frac{2}{(2 \pi)^{3/2}} \int_0^\infty dx\, 
    \frac{\tilde{g}_1^+(ix)}{x - i \omega} e^{-2Ax} 
    \exp \left[ \frac{ux}{\pi} \left( -1 + \ln\frac{ux}{\pi} \right) \right] \\
    & \times\operatorname{Im} 
    \left\{ \Gamma\left[ \frac12 - \frac{u(x-i\epsilon)}{\pi} \right] e^{iux} \right\},
\end{split}
\end{equation}
with $\epsilon\to 0$. The factor $\exp(-2Ax)$ suggests that the main contribution to the integral \eqref{Q_2_plus} comes from the region near $x=0$, leading to
\begin{equation}
    Q_c^+(\omega) \simeq 
    \frac{u}{\pi \sqrt{\pi e}}
    \left(
    C_i^{(\mathcal{R})} 
    + 2 \Theta^{(\mathcal{R})}\cos \frac{t \pi}{2u} 
    \right)
    \exp \left( -\frac{\pi A}{2u} \right) 
    \int_0^\infty dx\, \frac{x e^{-2Ax}}{x-i\omega}.
\end{equation}
Using \eq\eqref{WH_equation_factorization_2}, $\tilde{g}_c^+(\omega)$ is given by
\begin{equation}
    \tilde{g}_c^+(\omega) \simeq \frac{u}{\pi \sqrt{\pi e}} 
    \left(
    C_i^{(\mathcal{R})} 
    + 2 \Theta^{(\mathcal{R})}\cos \frac{t \pi}{2u} 
    \right)
    \exp \left(-\frac{\pi A}{2u} \right) G^+(\omega)
    \int_0^\infty dx\, \frac{e^{-2Ax}}{x - i\omega} x.
\end{equation}

\end{appendix}

\bibliographystyle{JHEP}
\bibliography{refs}

\end{document}